\documentclass{aastex701}
\usepackage{xcolor}
\usepackage{fancyvrb}
\usepackage{rotating}
\usepackage{booktabs} 
\usepackage{pgfplotstable}
\usepackage{subcaption}
\usepackage{longtable}

\pgfplotsset{compat=1.18}

\begin{document}

\title{White Dwarf Classification of DESI DR1 Spectra}

\author[orcid=0000-0002-8617-9317]{L. L. Amorim}
\affiliation{Instituto de F\'{\i}sica, Universidade Federal do Rio Grande do Sul, 91501-970 Porto Alegre, RS, Brazil}
\email[show]{larissal.pesquisa@gmail.com}

\author[orcid=0009-0000-3464-2026]{W.N. Costa Junior}
\affiliation{Instituto de F\'{\i}sica, Universidade Federal do Rio Grande do Sul, 91501-970 Porto Alegre, RS, Brazil}
\email{weligtonjunior1994@gmail.com}

\author[orcid=0000-0002-7470-5703]{S. O. Kepler}
\affiliation{Instituto de F\'{\i}sica, Universidade Federal do Rio Grande do Sul, 91501-970 Porto Alegre, RS, Brazil}
\email{kepler@if.ufrgs.br}

\author[orcid=0009-0002-2053-3026]{Jo\~ao Gabriel Leite Medeiros}
\affiliation{Instituto de F\'{\i}sica, Universidade Federal do Rio Grande do Sul, 91501-970 Porto Alegre, RS, Brazil}
\email{jgabriellm@protonmail.com}

\author[orcid=0000-0002-6164-6978]{Detlev Koester}
\affiliation{Institut f\"ur Theoretische Physik und Astrophysik, Universit\"at Kiel, 24098 Kiel, Germany}
\email{koester@astrophysik.uni-kiel.de}

\author[orcid=0000-0002-0797-0507]{Alejandra D. Romero}
\affiliation{Instituto de F\'{\i}sica, Universidade Federal do Rio Grande do Sul, 91501-970 Porto Alegre, RS, Brazil}
\email{aleromero82@gmail.com}

\begin{abstract}
We present a new catalog of spectroscopically confirmed white dwarfs from the Dark Energy Spectroscopic Instrument (DESI) Data Release 1. We visually classified 44\,417 white dwarf spectra and derived atmospheric parameters for 29\,072 DA white dwarfs through spectroscopic model fitting. The resulting mass distribution is non-Gaussian, with a mean mass of $0.677\,M_\odot$, consistent with previous studies. We identify 547 magnetic white dwarfs by detecting Zeeman splitting, including 84 new discoveries, and estimate their magnetic field strengths using off-centered, inclined dipole models when possible. We compare our magnetic field determinations with previous measurements and find overall good agreement. Finally, we investigate the relation between stellar properties and magnetism, finding that magnetic white dwarfs are systematically more massive than the general white dwarf population and that intermediate-strength magnetic fields are already present in stars that have not yet entered the crystallization phase. This result suggests that crystallization is unlikely to be the sole mechanism responsible for the origin of magnetic fields in white dwarfs.

\end{abstract}

\keywords{\uat{Stellar astronomy}{1583}, \uat{Stellar effective temperatures}{1597}, \uat{Stellar properties}{1624}, \uat{Stellar atmospheres}{1584}, \uat{Stellar types}{1634}, \uat{Stellar astronomy}{1583}, \uat{White dwarf stars}{1799}}

\section{Introduction}

When a star with an initial mass lower than approximately 10\,M$_\odot$, depending on its initial metallicity, runs out of fuel for fusion, it loses its outer layers, possibly forming a planetary nebula and a hot compact object at its center \citep{Fontaine2001}; we call it a White Dwarf. This includes at least 97\% of all stars \cite{lauffer2018}, which account for 6\% of stars in the solar neighborhood, and over 350\,000 have been identified with high confidence \cite[e.g.][]{Fusilo2021,blouin2024whitedwarffundamentals}. They are also possible outcomes of the evolution of multiple systems, with 25–30\% of white dwarfs estimated to result from mergers \citep{Toonen2017}.
To better understand the different groups of white dwarfs and their peculiarities, they are commonly classified into well-established spectral categories based on the dominant spectral species. Usually, the observed spectral features are absorption lines; however, emission lines may also be present, either chromospheric or when a companion acts as a donor star for the white dwarf. In all cases, each element or ion is characterized by the wavelength of the photons it emits or absorbs. 

The spectral classification is frequently done by examining the visible region of the spectra, where most surveys are centered and where lines from multiple elements of interest are observable, except for the highest temperatures and heavier elements, with lines visible only in the ultraviolet. In the optical region, for example, are the Balmer series of Hydrogen. If these lines are detected, the star is classified as a DA. Yet, for these features to appear, the gas must be at a temperature at which the electrons are changing levels. For Hydrogen, the lower limit is about 5000\,K, consistent with its ionization properties \citep{Alecian2015}. For Helium, another parameter is also at play: at temperatures close to 40\,000\,K, the atom is ionized, and if the observed lines are from He\,II, the white dwarf is classified as a DO. At lower temperatures, when the neutral element He\,I is dominant, the star is classified as a DB star if no Hydrogen lines are observed. At low enough temperatures, no spectral features are observed; once only the continuum is seen, they are classified as DC. On special occasions, a star may present no spectral features even at higher temperatures. We also classify them as DCs. 

Because gravitational settling operates on timescales much shorter than white dwarf cooling ages, most white dwarfs exhibit atmospheres dominated by either hydrogen or helium \citep{1979Fontaine,2008Fontaine}. Approximately 80\% of spectroscopically identified white dwarfs belong to the DA class \citep[e.g.][]{2019Kepler}. Even with this well-known gravitational sedimentation process, there are still white dwarfs with metals in their atmospheres. The fraction is estimated to be between 20 and 50\% \citep[e.g.][]{Hollands2018}. Three main mechanisms can ``pollute" their atmospheres: the presence of a deep convective layer that brings settled material back to the surface, gravitational levitation, and the ongoing accretion of external material such as planetesimals. Radiative levitation is only important at ($T_{eff} \geq 50\,000$~K), where it can significantly contribute \citep{1983Bruhweiler,1989Chayer,2014Barstow}. Although different in essence, both groups of stars are classified as DZ. We note that these suffixes are additive; a star with hydrogen, helium, and metal contamination would be called DABZ, with the order corresponding to the proportions. As discussed earlier, the spectral features can be in emission; this adds an `e' to the end of the acronym.

Although it is tempting to group all elements heavier than helium under the generic term ``metals'', two elements deserve special attention in white dwarf classification: carbon and oxygen. Carbon gives rise to the DQ spectral class, while oxygen can dominate the spectra of a small number of peculiar white dwarfs, classified as DS \citep{Williams19}. Cool white dwarfs whose spectra are dominated by carbon features often exhibit molecular absorption bands rather than sharp atomic lines. The appearance of these features depends strongly on the effective temperature, leading to the hot DQ and cool subclasses that reflect different atmospheric conditions and carbon-bearing species \citep{2005Dufour}. 

Finally, magnetic fields can also be identified in spectra through the Zeeman effect, which splits lines into multiple components by lifting the energy degeneracy of atomic energy levels. White dwarfs exhibiting magnetic fields are assigned the suffix H in their spectral classification. In cases of very strong magnetic fields, the stellar continuum may become measurably polarized, whereas very weak fields may be detectable only through the polarization signatures of spectral lines. When polarization is observed, the suffix P is added to the classification.

A group of stars that has historically been intertwined with white dwarf classifications is that of the subdwarfs (sd). Over time, this classification has become increasingly observational rather than strictly phenomenological. Several evolutionary pathways can produce stars with similar colors and luminosities. For example, (pre-)extremely low-mass white dwarfs (ELMs), blue stragglers, and metal-poor A/F stars may all be classified as subdwarfs based on their observed properties \citep{Heber16}. At optical wavelengths, white dwarfs can often be distinguished from sd stars through the Stark-broadened Balmer line profiles. The high surface gravities of DA white dwarfs produce significantly broader Balmer lines than those observed in sdA stars. The presence of metallic absorption features, such as the Ca II K line, also favors an sd classification. When available, atmospheric parameter determinations and Gaia-based luminosities provide a more robust distinction between the two classes, as subdwarfs are generally significantly more luminous than white dwarfs \cite{2018Pelisoli}.

Stars occupying the region near the white dwarfs in the HR diagram may also be binary systems, including double-degenerate systems (DDs), white dwarfs with cool red companions (DX+M), and interacting binaries in which a Roche-lobe-filling donor transfers material onto a white dwarf, known as cataclysmic variables (CVs).
Double-degenerate systems are often identified spectroscopically through poor fits to both the absorption lines and the continuum when modeled as a single star, indicating that the observed spectrum is instead a composite of two stellar components. A similar situation may occur for some apparent DAB stars, where the hydrogen and helium lines imply different effective temperatures, suggesting that the spectrum is produced by two distinct objects rather than a single mixed-atmosphere white dwarf.

When a spectrum covers a wide wavelength range, extending well into the red beyond the $H_\alpha$ line, this broad coverage enables the detection of cool companions, whose spectral features become more prominent at longer wavelengths. Because of their low temperatures, these companions typically exhibit broad molecular bands rather than discrete absorption lines. They can also be identified through spectral fitting, as the observed flux often exceeds that predicted by a single white dwarf model in the red portion of the spectrum, indicating the presence of a cool companion.

Cataclysmic variables are characterized by mass transfer onto the white dwarf. In the accretion regions, the gas is heated to sufficiently high temperatures to produce strong emission lines, which serve as its primary spectroscopic signature. Systems in which the accreted material is helium-rich and the spectrum is dominated by helium emission lines are classified as AM~CVn stars.

Studies of white dwarfs have expanded significantly, thanks to large-scale space telescope surveys such as Gaia Data Release 3, which has enabled the discovery of $\approx$ 354\,000 white dwarf candidates with high reliability \citep{Fusilo2021}. Over the last two decades, the Sloan Digital Sky Survey (SDSS) has helped to confirm and classify many of these objects \citep[e.g.][]{GF19, Kepler2015, Kepler2019, Kepler_2021}. From this perspective, the Dark Energy Spectroscopic Instrument (DESI,\cite{DESICollaboration2025}) can provide a new outlook on white dwarf studies, and with EDR \citep{Manser2024}, 2706 white dwarfs were identified spectroscopically.  

In this work, we investigate a newly released sample of spectra that became openly accessible with DESI data release DR1. In Section~\ref{cap:data}, we explain our selected sample, and in Section~\ref{cap:method}, we present our methodology for classifying these white dwarfs. In Sections~\ref{cap:results} and~\ref{cap:disc}, we present our results and interpret how our classification relates to the literature, as well as what other insights we can gain from such a large sample. 

\section{Data}
\label{cap:data}

\subsection{DESI}

The Dark Energy Spectroscopic Instrument (DESI) is a multi-object spectrograph capable of obtaining fiber spectra for approximately 5\,000 targets in a single pointing \cite{colaboration2022}, and it is mounted on the Mayall 4-meter telescope at Kitt Peak National Observatory (KPNO). Its fibers cover a wavelength range from 3600 to 9828~\AA, with a full width at half maximum (FWHM) resolution of about 1.8~\AA, and a spectral resolution of about 0.8~\AA \cite{colaboration2022}. DESI DR1 is the result of 13 months of observations, from May 2021 to June 2022. The resulting spectra are wavelength- and flux-calibrated through the DESI data processing and reduction pipeline (see \cite{colaborations2023} for details).

\subsubsection{DESI Spectra}

\begin{figure}[htbp]
    \centering
    \includegraphics[width=0.75\linewidth]{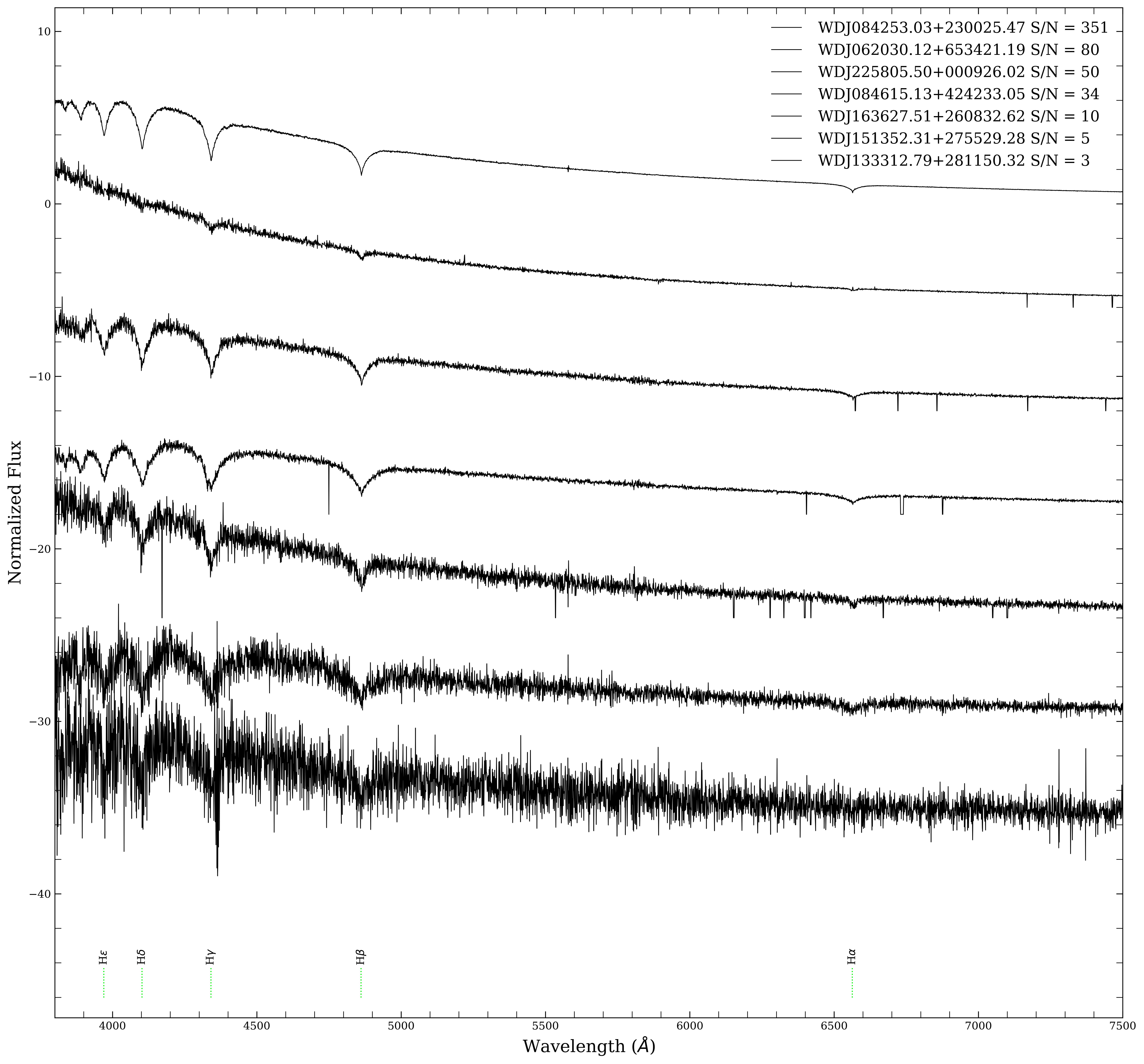}
    \caption{Spectra of seven DAs, showing different signal to noise ratio, from top to bottom, WDJ084253.03+230025.47, $S/N_B = 351$, MG = 9.9; WDJ062030.12+653421.19, $S/N_B = 80$, MG = 8.27;
    WDJ225805.50+000926.02, $S/N_B = 50$,  MG = 10.50; WDJ084615.13+424233.05, $S/N_B = 34$, MG = 11.3; WDJ163627.51+260832.62, $S/N_B = 351$, MG= 9.9; WDJ151352.31+275529.28,  $S/N_B = 5$, MG = 11.10 ; WDJ133312.79+281150.32, $S/N_B = 3$, MG = 10.50. 
    }
    \label{sn}
\end{figure}

The Milky Way project, a part of DESI,  contains over 4 million spectra, and we did not have the facilities to fit and identify every spectrum. 
Our selection criterion for the spectra was initially based on the work of \cite{Fusilo2021}, who created a catalog of over 1.2 million white dwarf candidates using Gaia DR3. We performed a cross-match between the DESI database and all stars in their catalog with a positive white-dwarf probability, using a matching radius of 1 arcsec, resulting in a total of 46\,006 targets. We found approximately 20\,000 targets with more than one observation in DESI, and we analyzed the spectrum with the largest S/N. Our analysis focuses on all objects with $S/N \geq 3$, which we adopt as the minimum threshold for reliably identifying H and He absorption features \citep[e.g.][]{dr7, Kepler_2021} for a total of 44\,417 targets. Figure \ref{sn} shows 7 spectra between $351 \leq S/N \leq 3$.

\begin{figure}[ht]
    \centering
        \includegraphics[width=\linewidth]{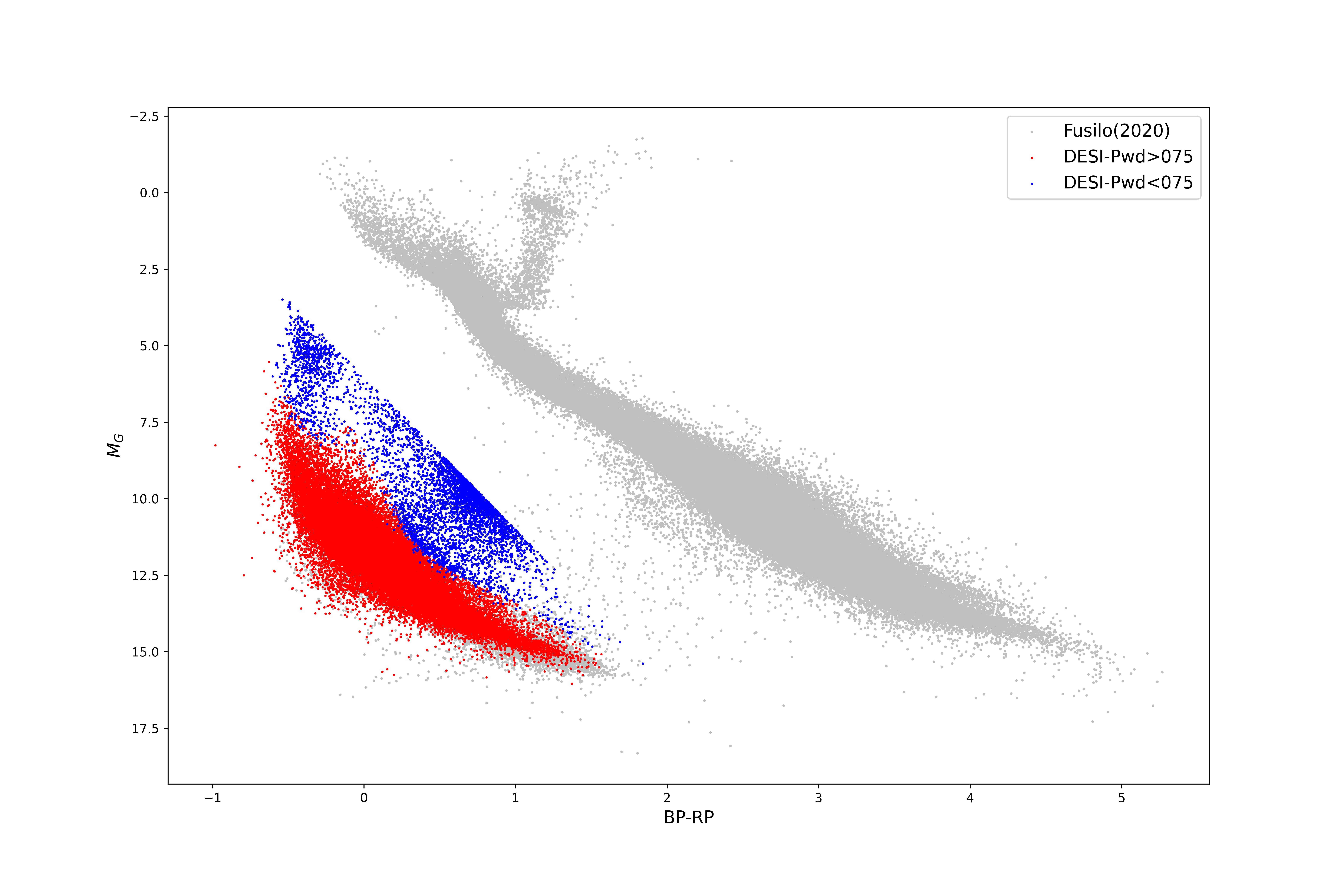}
        \caption{}
        \label{fig:HR}

    \caption{HR diagram with all targets, in red ($P_\mathrm{WD}\geq0.75$) and blue ($P_\mathrm{WD}<0.75$), from \cite{Fusilo2021}, and in grey the Gaia best sample to locate the main sequence.}

\end{figure}

We cross-matched our initial sample with previously published spectroscopic classifications compiled in the SIMBAD database and with catalogs from the SDSS and Large Sky Area Multi-Object Fiber Spectroscopic Telescope (LAMOST) surveys. This comparison showed that approximately 20\,660 objects had no spectroscopic classification available in the literature prior to the DESI observations. More recently, using DR1, \cite{2026Kilic} identified 7777 of these targets using the color criterion $G_{BP} - G_{RP} \leq 0$, which excludes more than 50\% of our initial sample. This work aims to include most targets in the sample with a high-confidence white-dwarf probability $PWD \geq 0.75$. With this, we now know that 11,685 targets matching \cite{Fusilo2021} are possibly being classified for the first time, as they were not found in any of the major white dwarf catalogs. 
The H-R diagram in Figure~\ref{fig:HR} shows our sample of white dwarf candidates from DR1, and under our selection criteria, 95.22\% of our targets have $PWD \geq 0.75$. Moreover, because DESI is a magnitude-limited survey ($M_G \sim 5$), intrinsically bright objects like subdwarfs and low-mass white dwarfs are selectively overrepresented in the data. 

\section{Methodology}
\label{cap:method}
\subsection{Classification}

\begin{figure}[htbp]

    \centering

    \begin{subfigure}{0.48\textwidth}
        \centering
        \includegraphics[width=\linewidth]{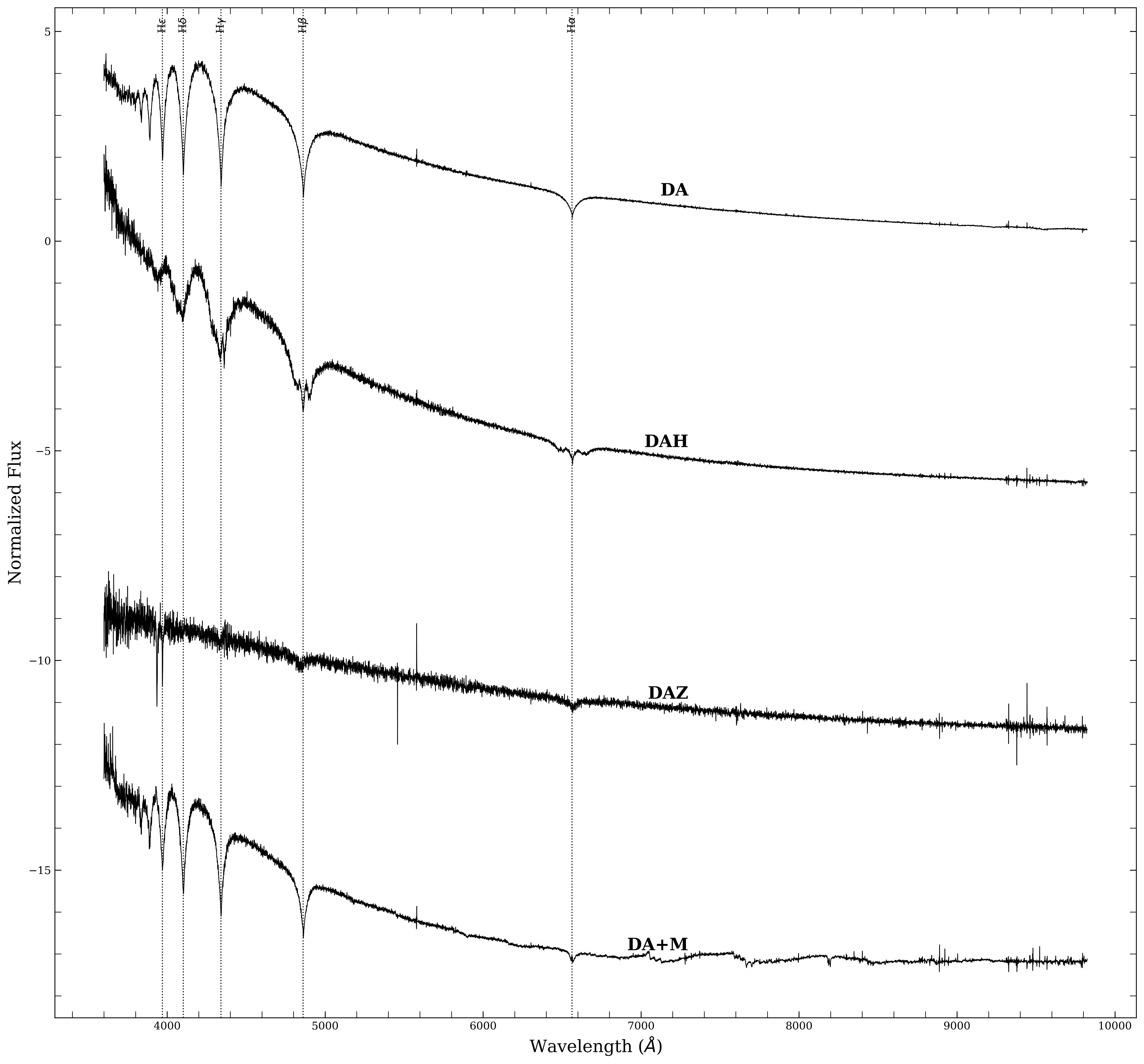}
        \caption{}
    \end{subfigure}
    \hfill
    \begin{subfigure}{0.48\textwidth}
        \centering
        \includegraphics[width=\linewidth]{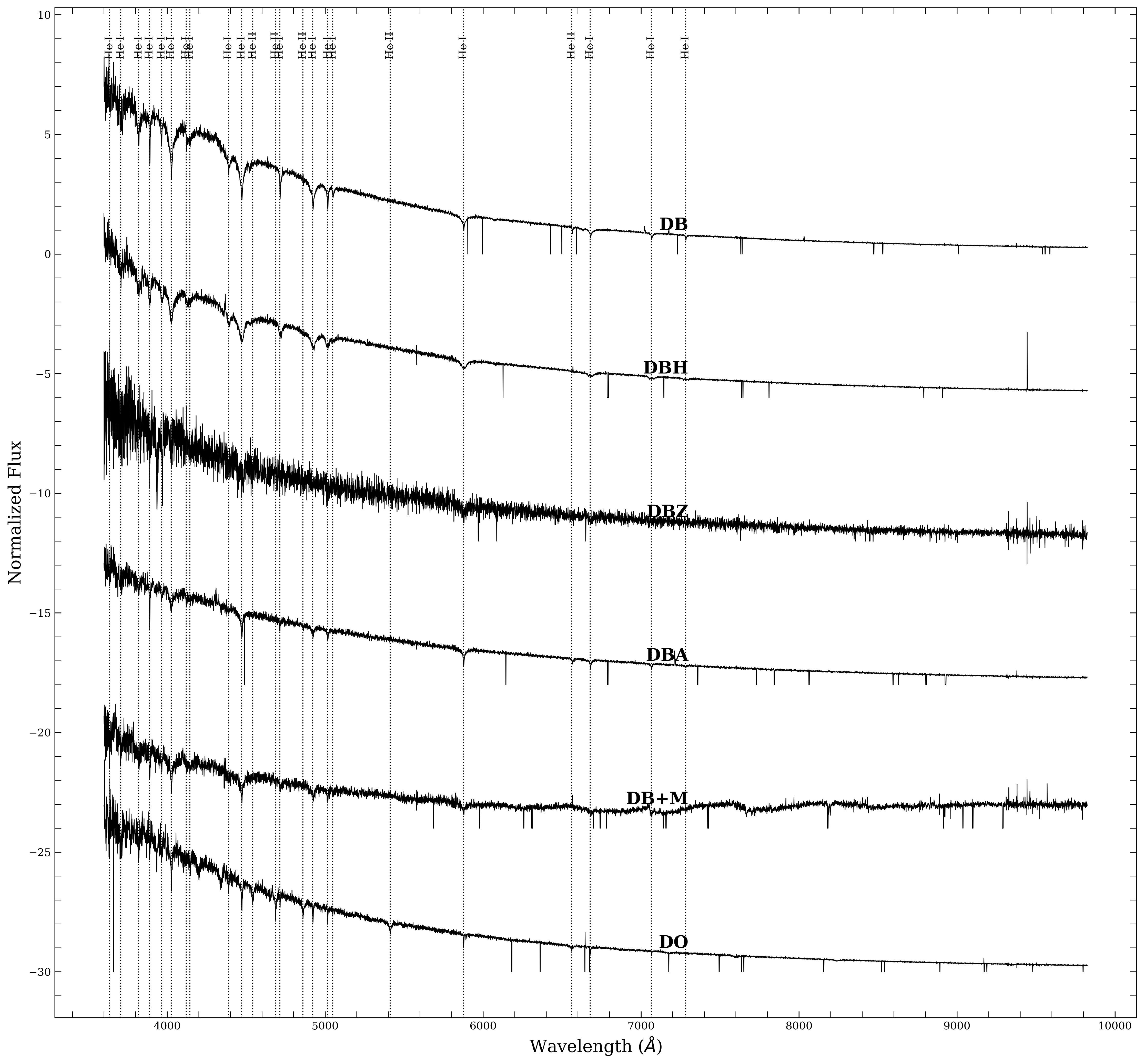}
        \caption{}
    \end{subfigure}

    \vspace{0.5cm}

    \begin{subfigure}{0.6\textwidth}
        \centering
        \includegraphics[width=\linewidth]{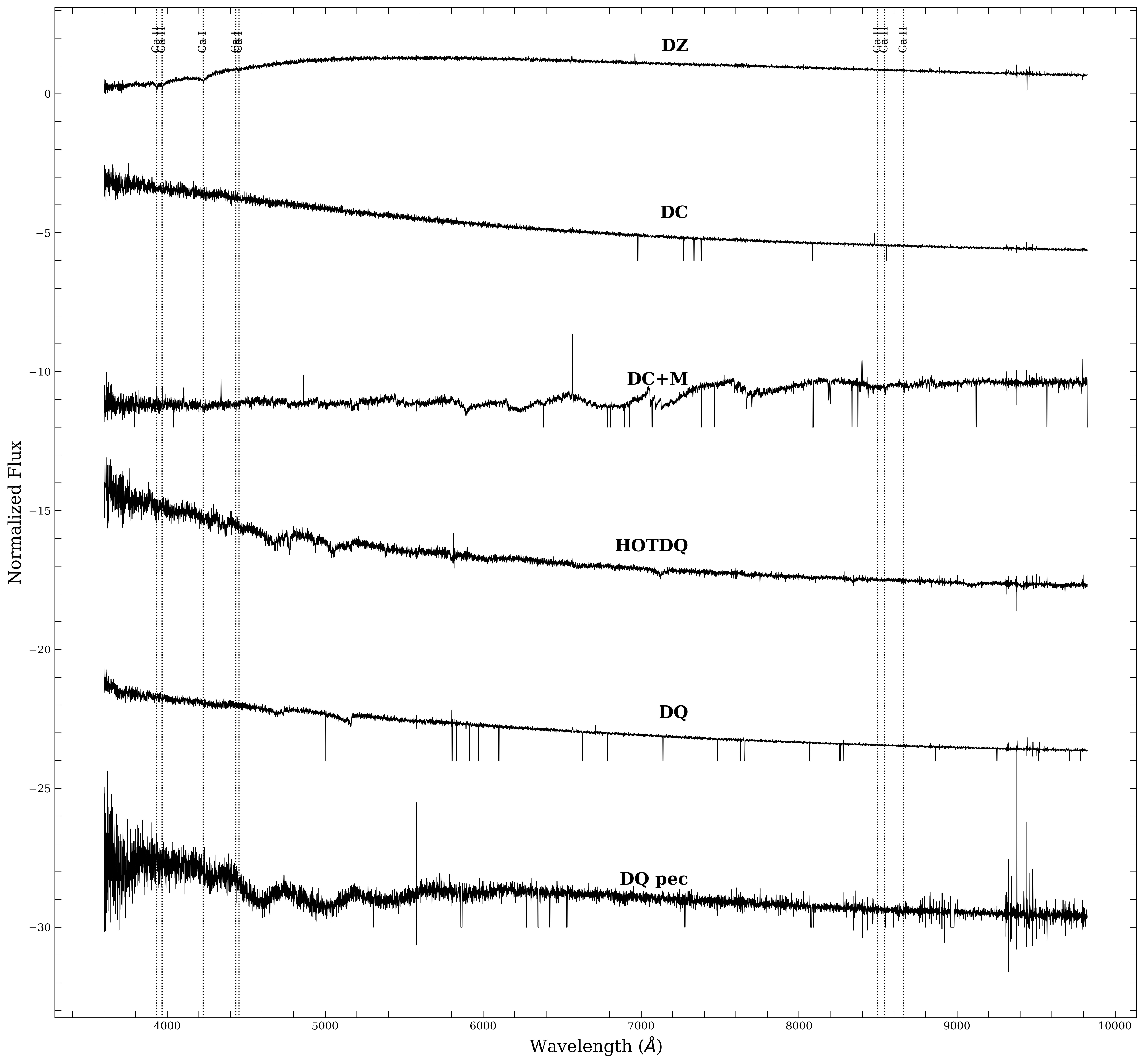}
        \caption{}
    \end{subfigure}
    
    \caption{Representative DESI spectra illustrating the main white dwarf spectral subclasses identified in this work. Panel (a) shows hydrogen-dominated white dwarfs (DA, DAH, DAZ, and DA+M), panel (b) helium-dominated white dwarfs (DB, DBH, DBZ, DBA, DB+M, and DO), and panel (c) other white dwarf subclasses (DZ, DC, DC+M, HotDQ, DQ, and DQ pec). The spectra are normalized and vertically offset for clarity. Vertical dashed lines mark the wavelengths of the principal diagnostic spectral features used for the spectral classification.}
    \label{fig:subclass}
\end{figure}

We visually inspected each spectrum individually, using the most prominent spectral features characteristic of each class as a guide. We used the features in Table~\ref{tab:lines} to assist in classifying each specified white dwarf subtype, see Figure \ref{fig:subclass}. A conservative classification scheme was intentionally adopted for the DA spectral class, motivated by the need to construct a robust sample for subsequent atmospheric and mass-distribution analyses. We assigned a spectrum to the pure DA category only when no detectable evidence of additional atmospheric species, magnetic fields (see section \ref{magfield}), or other peculiarities was present. Spectra displaying even subtle indications of such features were instead assigned to mixed or peculiar subclasses, with uncertainty indicators (:) added whenever the classification was not definitive. 

While some objects currently classified as mixed or uncertain subtypes may be reclassified as pure DA white dwarfs with future higher-quality observations, it is unlikely that a significant fraction of the clean DA sample will reveal previously undetected features within the sensitivity limits of the present data. Therefore, the resulting DA samples should provide a reliable basis for statistical investigations of white dwarf atmospheric properties and mass distributions. 

We found many objects with both strong Balmer lines and He\,I lines. These objects may be double-degenerate binaries composed of a DA and a DB white dwarf; however, since we did not fit multiple models, we followed standard nomenclature and simply classified them as DAB or DBA, as appropriate. For the 

We distinguished white dwarfs from subdwarfs only after fitting the atmospheric models, as the two classes can exhibit very similar optical spectra. Objects with best-fit surface gravities of $\log\, g<7$ were classified as sdA or sdB, depending on whether their effective temperatures were below or above approximately 18\,000~K, respectively. Since this temperature does not represent a sharp boundary between the two classes, we used Gaia-based effective temperature estimates to validate the classification. In cases where the spectroscopic and Gaia-based temperatures disagreed, and the estimated temperature lay near the transition region (approximately 16000--20000~K), we adopted the uncertain classification sdA/sdB. When no reliable effective temperature estimate was available, we assigned the more general classification sd:.

When spectra showed no detectable absorption or emission lines, we classified them as DC white dwarfs. Since weak spectral features can become undetectable at low $S/N$, the DC classification was assigned conservatively, requiring the absence of identifiable diagnostic lines across the observed wavelength range. Objects displaying weak continuum contamination from unresolved companions were classified separately as DC+M. 

Objects exhibiting unusual carbon features or spectral morphologies that did not match the standard DQ sequence were classified as DQ pec. The distinction between DQ, HotDQ, and DQ pec follows the traditional white dwarf spectral classification scheme and is based on the dominant carbon features in the optical spectrum, with the majority exhibiting the characteristic C$_2$ Swan molecular bands.

Subsequently, we compared our classification with that in \cite{2026Kilic}. For the mismatched cases, we further investigate the possible origin of the discrepancy and present our findings in Section~\ref{cap:results}.

\begin{table}
\tabletypesize{\scriptsize}
\centering
\begin{tabular}{c|c|c}
\hline
\textbf{Spectral Type} & \textbf{Diagnostic Feature} & \textbf{Main Lines/Bands (\AA)} \\
\hline

DA &
Broad Balmer absorption lines& H$\eta$ 3835.38, H$\zeta$ 3889.05, H$\epsilon$ 3970.07, \\
 & with a pronounced Balmer decrement &

H$\delta$ 4101.74, H$\gamma$ 4340.47,
H$\beta$ 4861.33, H$\alpha$ 6562.80 \\
\hline

DB / sdB &
Neutral helium (He\,I) absorption lines &
3964.73, 4026.19, 4120.82, 4143.76, 4387.93,\\
 & &{\bf 4471.48}, 4713.15, 4921.93, 5015.68,\\
 & &5047.74, 5875.70, 6678.15, 7065.19, 7281.35 \\
\hline

DO / sdO &Ionized helium (He\,II) absorption lines &
4541.59, {\bf 4685.68}, 4859.32, 5411.50, 6560.10\\ 
PG1159 && \\
\hline

DQ &
Atomic carbon lines and/or C$_2$ Swan bands &
C I features and molecular C$_2$ bands (typically 4380--6200) \\
\hline
Hot DQ &
Ionized carbon (C II) absorption lines &
C\,II 4367 and associated carbon features \\
\hline

DZ & Metal lines, particularly Ca II H\&K&Ca\,II H (3968.47), Ca\,II K (3933.66);\\
DAZ / DBZ & &
 often Mg\,I and Fe lines \\
\hline

DH &
Zeeman splitting caused by a magnetic field &
Split hydrogen, helium, or metal lines \\
\hline

DC &
Featureless continuum spectrum &
No prominent spectral features \\
\hline

CV &
Strong emission-line spectrum &
Balmer and He emission lines \\
\hline

\end{tabular}
\caption{Main spectral features used in the classification of white dwarfs and related compact stellar objects.}
\label{tab:lines}
\end{table}

\subsection{Effective Temperature and $\log g$ fits}

For DAs, we fitted the spectra to synthetic models to determine the effective temperature and log g. 
We did not fit composite white dwarf plus main-sequence star models to the apparently composite spectra, as done by \cite{RM09,RM21,RM25,RM26,2025Morales}. Instead, a dedicated study focusing on binaries with cool companions and detailed spectral modeling is currently submitted (Morales et al. 2026).

We fitted synthetic spectra calculated by Detlev Koester for the DA sample using $\chi^2$ minimization. These theoretical spectra were calculated assuming LTE and are updated versions of \cite{Koester10}. Our model grid covers an effective temperature range of 6,000 --- 80,000~K and a $\log g$ range of 7.0 --- 9.5. These intervals encompass the atmospheric parameters of the vast majority of known DA white dwarfs \citep{2019Kepler}. Unfortunately, in a few rare cases, the model that best fits the data lies at the edge of the adopted grid, suggesting that an even better solution may exist outside the explored parameter space. For these cases, we inspected the fit to verify whether the stars really have high/low effective temperature/log g, or whether another factor may have misled the fitting code.

Most of the stars analyzed in this way turned out to be genuine DAs for which our fitting algorithm did not find an adequate solution due to insufficient signal-to-noise ratio, too many bad pixels in the spectra, or uncertain colors. Many were also multi-subtype white dwarfs, such as DAH, DAB, DAZ, or DA+M.
Considering all fits, a group of objects showed Balmer lines shallower than expected for DA white dwarfs at their derived effective temperatures. These stars are most likely double degenerates consisting of a DA and a DC white dwarf, but we kept them as DAs.

We finally note that the white dwarf color space also contains many hot subdwarfs. It is difficult, just by looking at a spectrum, to tell a low-mass white dwarf from a subdwarf. To guide us in classifying hot stars, we superimposed a $\log{g}=7$ model on the observed spectra, using the effective temperature from a fit and/or Gaia photometry.  We then rejected objects showing lines much narrower than the synthetic spectrum.  

The DAWD\_Fitting\footnote{\url{https://github.com/Mandorama/DAWD\_Fitting}} code was used to determine the best-fitting model. It performs linear interpolation between models to improve the agreement with the observed data. Model comparison and selection are performed using the chi-square ($\chi^2$) minimization. 
After obtaining the effective temperature and surface gravity, one can determine the stellar mass using evolutionary models. We used models from \cite{2013Althaus} and \cite{lauffer2018} for stars with Carbon-Oxygen and Oxygen-Neon cores, respectively.

\subsection{Magnetic field intensities}
\label{magfield}
For stars classified as magnetic and exhibiting hydrogen-dominated spectra, we applied off-centered inclined dipole models to fit the observations and derive magnetic field strengths. This approach has been widely adopted because it accounts for the continuous distribution of magnetic field intensities across the stellar surface, resulting in a more realistic representation of the magnetic field geometry and generally providing better agreement with observed spectra. We used the YAWP code developed by \cite{2009Kulebi} for this purpose.
The model grid implemented in YAWP is limited to effective temperatures above $T_{eff} = 8000\,K$. Therefore, for cooler stars, we were unable to determine reliable magnetic field strengths. These objects were consequently classified simply as DAH stars, without a quantitative estimate of their magnetic field intensity. The models are also limited to $\log\,g = 8$.
A comparison between the results of \cite{2023Hardy} and \cite{2023amorim} shows that the field intensity is generally unaffected by this limitation.

For stars previously identified as magnetic and with magnetic field strengths already reported in the literature, we adopted the published values. The two studies providing such measurements for DESI spectra are \cite{2026Kilic} and \cite{2026amorim}. We also investigated magnetic white dwarfs (MWDs) identified through other spectroscopic surveys, such as SDSS \cite[e.g.,][]{2023amorim,2025moss}, that were subsequently observed by DESI.

It is worth noting that some of these objects were already included in our sample because their magnetic fields were independently identified during our visual inspection. For stars selected solely from the literature, however, the YAWP fitting procedure did not converge to a satisfactory solution; they are marked in the final results as DNC. This failure does not imply the absence of a magnetic field. In most cases, these DESI spectra have low signal-to-noise ratios, and the reported magnetic field strengths are relatively weak, making reliable determination difficult with the available data.

\section{Results}
\label{cap:results}

The summary of our findings is available in Table~\ref{tab:resumo}. In Table~\ref{tab:high_velocity}, ,  together with the \href{https://github.com/AstroWeljr/DESI-DR1-Class}{DESI-DR1-Class repository}, we present our full classification list, including effective temperatures, surface gravity, and masses calculated in this work.

\begin{figure}[h]
    \centering
    \includegraphics[width=0.75\linewidth]{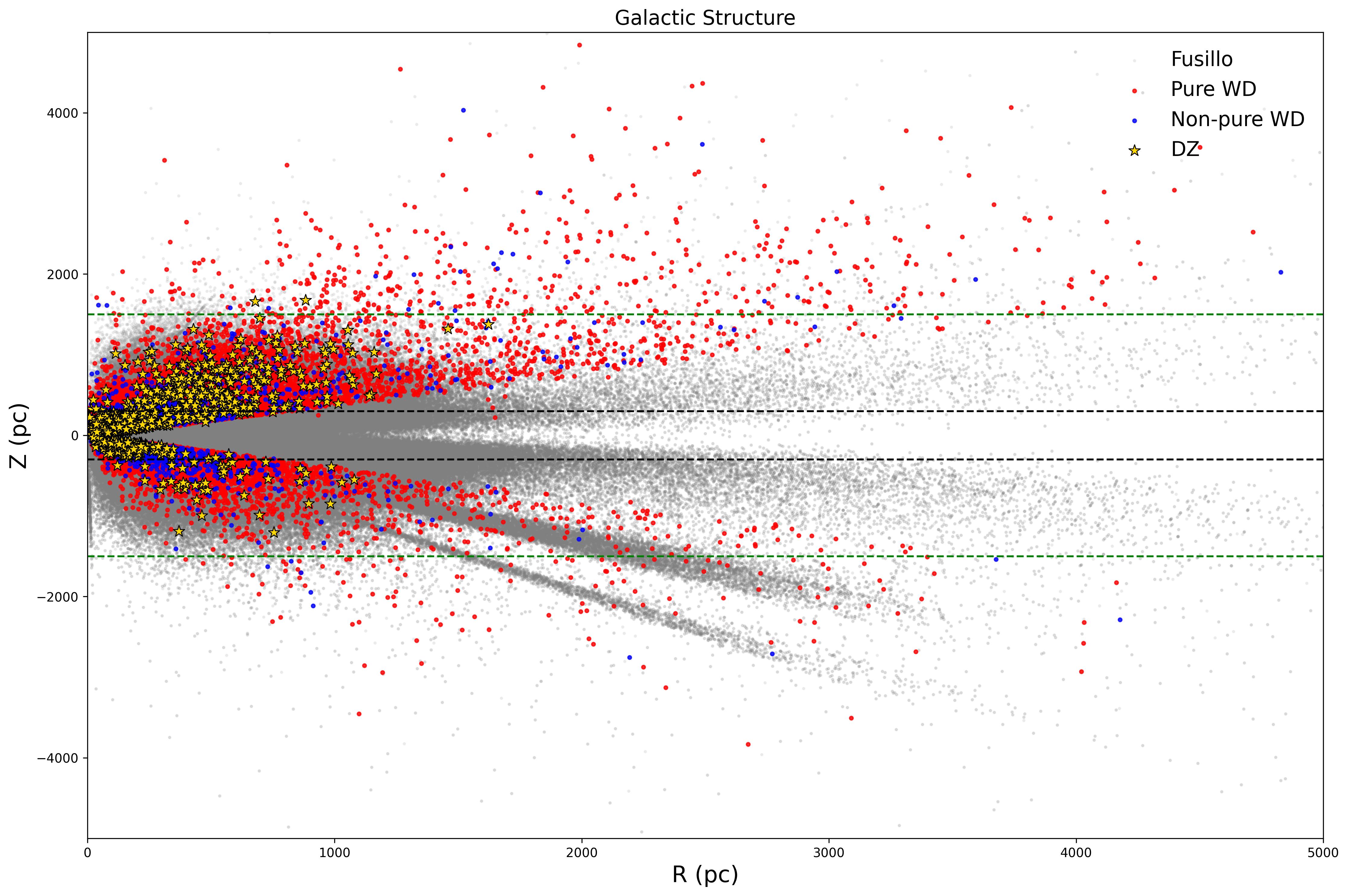}
    \caption{Galactic R-Z distribution of the white dwarfs classified in this work. Grey points correspond to the white dwarf sample from \cite{Fusilo2021}. Pure white dwarfs are shown as red circles, non-pure white dwarfs as blue circles, and DZ white dwarfs (including all DZ subclasses) as yellow star symbols. The black dashed lines indicate the approximate scale height of the Galactic thin disk ($|Z|\approx300pc$), whereas the green dashed lines indicate the approximate scale height of the thick disk ($|Z|\approx1500pc$).}
    \label{fig:galat}
\end{figure}

Figure~\ref{fig:galat} presents the spatial distribution of the white dwarfs in our sample, distinguishing pure white dwarfs from those exhibiting atmospheric metal pollution. The overall Galactic distributions of the two populations are similar, with both concentrated toward the Galactic plane. However, the metal-polluted white dwarfs are almost exclusively found within the nearby concentration of objects, whereas pure white dwarfs are detected over a much larger volume. The apparent absence of metal-polluted white dwarfs at larger distances is most likely a consequence of observational selection effects, since the weak metal absorption features that define these objects become increasingly difficult to detect in spectra with lower signal-to-noise ratios.

We searched the literature for previous spectral classifications using the catalogs of \cite{2024Vincent} and \cite{2026Kilic}, the Montreal White Dwarf Database (MWDD), and SIMBAD. Objects for which no previous spectral classifications were found in any of these resources are flagged as candidate new spectral classifications presented in this work. All of these objects have Gaia colors $(G_{\rm BP}-G_{\rm RP}) > 0$, placing them in the redder region of color space that was excluded from the analysis of \cite{2026Kilic}.

\begin{table}[h]
    \centering
    \begin{tabular}{ccc}
    \hline
        SpType	&	Complete sample	&	Possible new classification\\	\hline
        DA     	&	31512+5:	&	8419+0:	\\
        DB     	&	2598+13:	&	362+3:	\\
        DO     	&	168+3:	&	16+3:	\\
        DC     	&	3932+0:	&	1913+0:	\\
        DZ     	&	1021+25:	&	595+13:	\\
        DQ     	&	250+98:	&	111+47:	\\
        DS     	&	3+2:	&	1+0:	\\
        MWD    	&	547+218:	&	105+84:	\\
        Fe     	&	0+1:	&	0+1:	\\
        DX     	&	15+0:	&	6+0:	\\
        DXY    	&	581+5:	&	188+2:	\\
        DX+M   	&	805+2:	&	431+1:	\\
        DXe    	&	208+0:	&	65+0:	\\
        DD     	&	12+0:	&	0+0:	\\
        CV     	&	219+1:	&	156+1:	\\
        sd     	&	2134+39:	&	1221+17:	\\\hline
    \end{tabular}
    \caption{Number of white dwarfs classified in each spectral group (middle column) and the number of objects without previous spectroscopic classifications in the literature (right column). Objects with uncertain classifications are listed separately and denoted by a trailing colon (``:").}
    \label{tab:resumo}
\end{table}

\begin{figure}[h]
    \centering
    \includegraphics[width=0.8\linewidth]{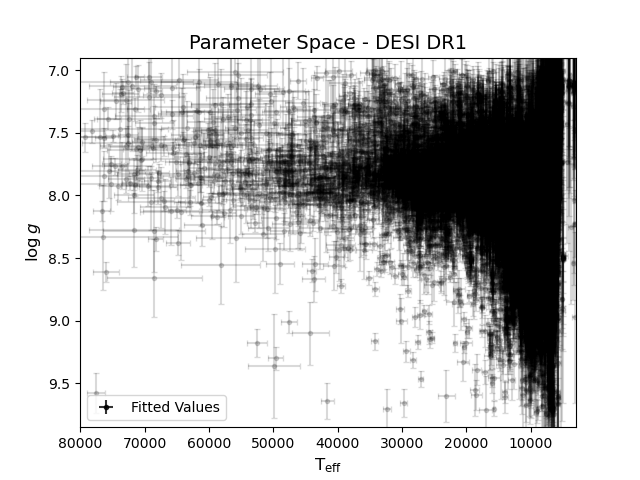}
    \caption{Stellar parameters determined for the white dwarfs successfully modeled.}
    \label{fig:parspace}
\end{figure}

\begin{sidewaystable}[h]
\centering
\caption{Candidate high-velocity white dwarfs. Full version available online}
\begin{tabular}{l l r r r r r r r r}
\hline
TIC & Name & RA (deg) & Dec (deg) & $V_\mathrm{RAD}$ (km s$^{-1}$) & $\varpi$ (mas) & PWD & G & BP-RP & $M_G$\\
\hline
671790783  & WDJ040338.30$-$104945.19 & 60.910 & -10.829 & -1499.35 & 3.083 & 0.96 & 18.694 & 0.150 & 11.139 \\
630446401  & WDJ014203.50+031202.42  & 25.515 & 3.200   & -1499.96 & 4.802 & 0.98 & 18.858 & 0.270 & 12.265 \\
620222514  & WDJ020633.46+205707.65  & 31.639 & 20.952  & 1499.81  & 1.858 & 0.02 & 18.172 & 0.659 & 9.518  \\
620164711  & WDJ022422.46+211327.79  & 36.094 & 21.224  & 1499.74  & 1.635 & 0.05 & 19.796 & 0.807 & 10.863 \\
1100207634 & WDJ151415.68+074446.75  & 228.565 & 7.746  & -1500.00 & 5.531 & 0.78 & 18.665 & 0.552 & 12.379 \\
1307383270 & WDJ170147.24+054540.48  & 255.447 & 5.761  & -1499.99 & 2.002 & 0.45 & 20.568 & 0.710 & 12.075 \\
1200186232 & WDJ160335.91+215032.56  & 240.900 & 21.842 & -1499.39 & 3.228 & 0.96 & 19.119 & 0.198 & 11.664 \\
2052592071 & WDJ225237.05$-$051916.97 & 343.154 & -5.321 & -65.15 & 1.953 & 0.92 & 19.041 & 0.077 & 10.494 \\
1002802815 & WDJ133540.76+050722.74  & 203.920 & 5.123 & 1153.12 & 0.454 & 0.25 & 19.009 & 0.147 & 7.295 \\
469348568  & WDJ092614.31+010557.26  & 141.560 & 1.099 & -1342.01 & 2.682 & 0.48 & 19.470 & 0.496 & 11.613 \\
1200064785 & WDJ160816.93+180525.33  & 242.071 & 18.090 & 1000.59 & 1.959 & 0.45 & 19.949 & 0.485 & 11.410 \\
2025118548 & WDJ214140.45+050730.24  & 325.418 & 5.125 & 15.49 & 3.713 & 0.99 & 19.002 & 0.062 & 11.851 \\
291157234  & WDJ124325.92+025547.37  & 190.858 & 2.930 & -1499.59 & 1.657 & 0.03 & 17.483 & 0.456 & 8.580 \\
802300170  & WDJ080701.10+554738.84  & 121.755 & 55.794 & 1500.00 & 1.614 & 0.15 & 19.645 & 0.559 & 10.685 \\
800454415  & WDJ085232.03+122735.07  & 133.133 & 12.460 & 1411.09 & 7.151 & 0.99 & 19.131 & 0.422 & 13.403 \\
952753275  & WDJ120619.87$-$032838.27 & 181.583 & -3.477 & -1500.00 & 1.483 & 0.39 & 20.178 & 0.490 & 11.033 \\
\hline
\end{tabular}
\label{tab:high_velocity}
\end{sidewaystable}
In Fig.~\ref{fig:parspace}, we present the distribution of the atmospheric parameters derived for our sample. A prominent overdensity of cool white dwarfs is apparent. This is expected from white dwarf cooling theory, as stars spend progressively longer times at lower temperatures; consequently, cooler white dwarfs are intrinsically more numerous in magnitude-limited samples.

The relative scarcity of hot, massive white dwarfs is also a consequence of their cooling evolution. More massive white dwarfs cool more rapidly at high effective temperatures due to their smaller heat capacities and enhanced neutrino losses, thereby spending less time in this region of parameter space. As a result, they are observed less frequently than their lower-mass counterparts at similar temperatures.

\section{Discussion}
\label{cap:disc}
In addition to producing the catalog itself, we report here on the increased number of magnetic white dwarf stars found in this catalog compared with previous catalogs. We also examine the mass distribution of our DA sample and find it to be non-Gaussian, as in previous studies \citep{2019Kepler}.

\subsection{Comparison with literature classification.}

Of the 19\,319 white dwarfs classified by \cite{2026Kilic}, 18\,961 are present in our sample. We disagree on the classification of only 1\,488 objects. Most of these discrepancies reflect different assessments of the significance of a secondary spectral component. Additional differences arise when \cite{2026Kilic} does not report a cool companion or emission lines that we identify, or when they classify an object as a double-degenerate (DD) system, a class that was not systematically investigated in this work.

\subsection{Effective Temperature Spectroscopic vs. Photometric}
Several independent determinations of the effective temperature are available for a subset of our targets. However, only a handful of studies have provided homogeneous determinations of atmospheric parameters for large samples of white dwarfs.



Since our sample was selected from the white dwarf catalog of \cite{GF19}, almost all objects have photometric temperature estimates based on Gaia data. Subsequently, Gaia released low-resolution $BP - RP$ spectra for an unprecedented fraction of the sky. Using these data, \cite{2024Vincent} performed spectral classifications and derived atmospheric parameters, including effective temperatures, for a large sample of white dwarfs. Unfortunately, fewer than 9\,000 are in the intersection with our sample.

\begin{figure}[h]
    \centering
    \includegraphics[width=0.45\linewidth]{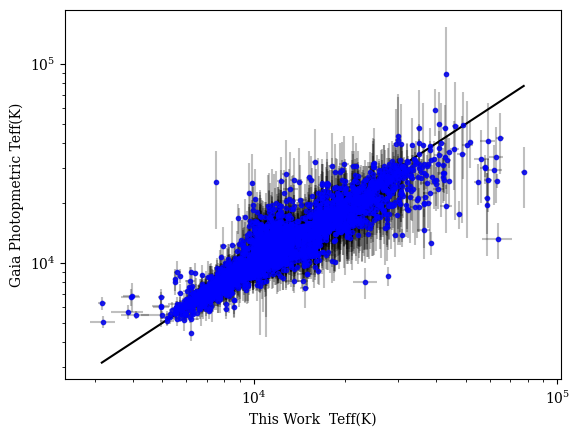}
    \includegraphics[width=0.45\linewidth]{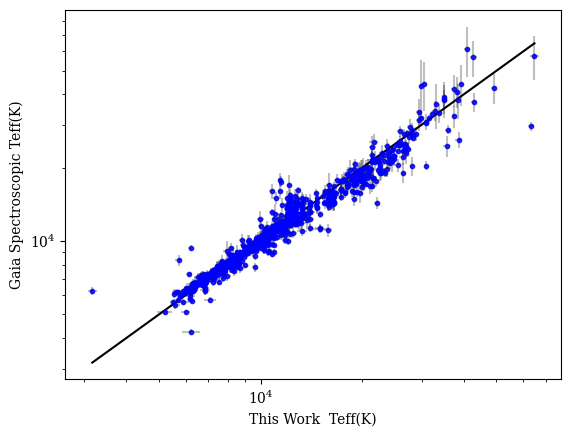}
    \caption{Comparison between effective temperatures derived by different methods.}
    \label{fig:teff}
\end{figure}
In Fig.~\ref{fig:teff}, we compare the effective temperatures derived in this work with those obtained from Gaia data using both photometric and low-resolution spectroscopic methods. Two features are particularly noteworthy. First, there is an overdensity of objects in the temperature range where the Balmer lines reach their maximum strength, producing the well-known degeneracy between effective temperature and surface gravity in spectroscopic fits. The second is the systematic difference between photometric and spectroscopic temperature estimates, a discrepancy that has been extensively documented in the white dwarf literature.
\subsection{Mass distribution}
\begin{figure}[h]
    \centering
    \includegraphics[width=0.5\linewidth]{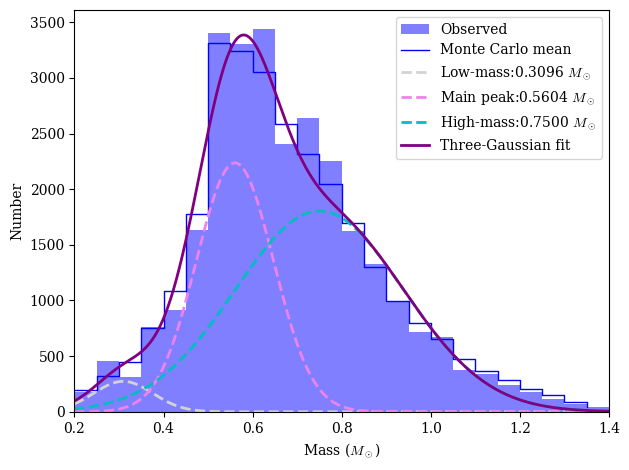}
    \caption{Histogram of the masses of the DA white dwarfs modeled in this work. The light blue histogram shows the observed mass distribution. The blue curve represents the mass distribution obtained from 500 Monte Carlo realizations, accounting for the individual measurement uncertainties. The dark purple curve shows the sum of three Gaussian distributions, illustrating different possible white dwarf populations.} 
    \label{fig:mass}
\end{figure}

Figure~\ref{fig:mass} shows the mass distribution of the DA white dwarfs for which we derived masses from evolutionary models, with a mean of $0.677\,M_\odot$ and a median $0.647\,M_\odot$. The distribution is clearly non-Gaussian: the primary peak lies below the mean mass, and a secondary peak is present at higher masses. This behavior has been extensively reported in the literature and is a well-established characteristic of large white dwarf samples. The high-mass component has been attributed to a combination of evolutionary channels, including binary evolution and white dwarf mergers, although its exact origin remains under debate.

To characterize the observed distribution, and following \cite{2007Kepler,Kepler2015}, we fitted a model consisting of three Gaussian components, with initial parameters chosen from the three prominent peaks of the histogram. While these Gaussian components do not necessarily correspond to distinct physical populations, they provide a convenient parameterization of the distribution and are broadly consistent with different evolutionary channels. In particular, the low-mass component is expected to be dominated by white dwarfs formed through binary interactions, whereas the high-mass component is thought to contain a significant contribution from merger products. Finally, to account for the uncertainties in the individual mass determinations, we generated 500 Monte Carlo realizations and used them to estimate the most probable underlying mass distribution.

\subsection{Magnetic Fields and Zeeman Splittings}

We present a compilation of all magnetic white dwarfs (MWDs) observed by DESI for which magnetic field strength determinations are available. In this work, we successfully modeled 210 DAH stars, of which 85 are newly identified magnetic white dwarfs. From \cite{2026amorim}, we include the 91 newly discovered DAH stars reported in that study. From \cite{2026Kilic}, we include 246 DAH stars, comprising both previously known and newly identified objects.

In addition, we report 74 DAH: stars corresponding to objects visually classified as magnetic, but for which no satisfactory magnetic field model could be obtained. These stars are likely magnetic white dwarfs, although a quantitative determination of their magnetic field strengths was not possible from the available data. Table~\ref{tab:mwd} displays the magnetic parameters.

\begin{table*}
\centering
\scriptsize

\begin{minipage}[t]{0.49\textwidth}
\centering
\begin{tabular}{cccc}
    \hline
    DESI ID & $B_{\rm lit}$ & Ref & $B_{\rm DESI}$ \\
         & (MG) &        & (MG) \\
    \hline
    39633103550940536	&	-	&	[3]	&	0.4$^*$	\\
    39628509085176751	&	-	&	[3]	&	0.7$^*$	\\
    39632934751175691	&	-	&	[2]	&	0.75$^\dagger$\\
    39628167417168076	&	1.4	&	[8]	&	0.8$^*$	\\
    39627725186532444	&	-	&	[3]	&	0.8$^*$	\\
    39628290180253631	&	-	&	[3]	&	0.8$^*$	\\
    39633365913043597	&	-	&	[3]	&	0.8$^*$	\\
    39628499304058947	&	-	&	[3]	&	0.9$^*$	\\
    39628509462660457	&	-	&	[3]	&	0.9$^*$	\\
    39627684665363317	&	2	&	[10]	&	1$^*$	\\
    39627586799666838	&	0.8	&	[9]	&	1.5		\\
    39633236430687287	&	1.0	&	[5]	&	1.5		\\
    39628174711064898	&	1.5	&	[6]	&	1.5$^*$	\\
    39628129546797836	&	1.9	&	[5]	&	1.5		\\
    39628204683559444	&	2.62	&	[1]	&	1.5		\\
    39628198891229530	&	-	&	[2]	&	1.5$^\dagger$\\
    39628270412501473	&	-	&	[3]	&	1.7$^*$	\\
    39627681456720752	&	-	&	[3]	&	1.8$^*$	\\
    39632991785322454	&	2.64	&	[1]	&	1.9$^*$	\\
    39628516555230071	&	-	&	[2]	&	1.92$^\dagger$\\
    39633066662039559	&	2.44	&	[1]	&	1.98		\\
    39632955101938456	&	2.46	&	[1]	&	2		\\
    2305843017975602293	&	2.49	&	[1]	&	2		\\
    39628112190767639	&	2.9	&	[5]	&	2		\\
    39633063918963763	&	8.0	&	[5]	&	2		\\
\end{tabular}
\end{minipage}
\hfill
\begin{minipage}[t]{0.49\textwidth}
\centering
\begin{tabular}{cccc}
\hline
DESI ID & $B_{\rm lit}$ & Ref & $B_{\rm DESI}$ \\
         & (MG) &        & (MG) \\
\hline
39627767788082401	&	-	&	[2]	&	2.0$^\dagger$\\
39633271843195931	&	-	&	[2]	&	2.0$^\dagger$\\
39632935967525271	&	2.36	&	[1]	&	2.02		\\
39633248355093325	&	2.29	&	[1]	&	2.03		\\
39627775614650145	&	2.0	&	[1]	&	2.04		\\
39627806841246365	&	2.16	&	[1]	&	2.05		\\
39628004233577324	&	2.64	&	[1]	&	2.05		\\
39633178448627254	&	2.83	&	[1]	&	2.05		\\
39633437438509791	&	-	&	[2]	&	2.05$^\dagger$\\
39633245008040061	&	-	&	[3]	&	2.1$^*$	\\
39627818581103370	&	2.72	&	[1]	&	2.16		\\
39633365241954558	&	-	&	[2]	&	2.20$^\dagger$\\
39633429788100776	&	-	&	$\rightarrow$	&	2.22		\\
39628054355511034	&	2.82	&	[1]	&	2.23		\\
39628100685797752	&	-	&	[2]	&	2.23$^\dagger$\\
39633342064231252	&	2.01	&	[1]	&	2.25		\\
39633328634072951	&	0.2	&	[10]	&	2.27		\\
39627812801351434	&	2.62	&	[1]	&	2.27		\\
39627565039616142	&	-	&	[2]	&	2.27$^\dagger$\\
39627946989717673	&	2.07	&	[1]	&	2.28		\\
39628260887234969	&	2.72	&	[1]	&	2.28		\\
39628291304331375	&	-	&	$\rightarrow$	&	2.28		\\
39627814059647373	&	-	&	[2]	&	2.28$^\dagger$\\
39627825136798655	&	-	&	[2]	&	2.28$^\dagger$\\
39628492119216740	&	-	&	[2]	&	2.28$^\dagger$\\
\end{tabular}
\end{minipage}

\caption{Magnetic white dwarfs with measured magnetic field strengths.[1]\cite{2023amorim},[2]\cite{2026amorim},[3]\cite{2026Kilic},[4]\cite{2023Hardy},[5]\cite{2025moss},[6]\cite{2023Hardy}2, [7]\cite{MWDD}, [8]\cite{2026AYu},[9]\cite{2022Bagnulo},[10]\cite{2024Jewett}.The strength measured from DESI spectra has three sources: \cite{2026Kilic}, $\dagger$\cite{2026amorim}, and those without marks are from this work. The newly identified DAHs are marked with an arrow. The full table is available in machine-readable form.}
\label{tab:mwd}
\end{table*}
We compared the magnetic field strengths derived from the DESI spectra with previously published measurements for the 164 stars common to both samples. The results show overall agreement: 78 stars ($\sim$48\%) differ by less than 1 MG, while 45 stars ($\sim$27\%) differ by less than 10\% relative to the literature values.

\begin{figure}
    \centering
    \includegraphics[width=0.5\linewidth]{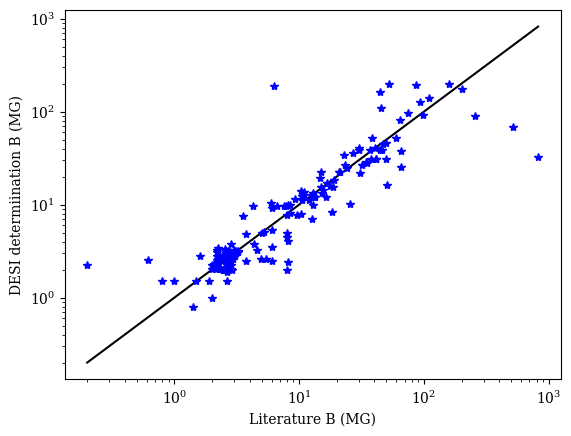}
    \caption{Comparison between magnetic field intensities derived from DESI versus from other surveys.}
    \label{fig:BxB}
\end{figure}

Although these numbers may initially appear modest, determining magnetic field strengths from spectral fitting is a highly degenerate problem, with multiple field geometries and atmospheric parameters often producing similar spectral features. Consequently, exact agreement between independent analyzes is not always expected. Furthermore, when studying the origin and evolution of magnetic fields in white dwarfs, the order of magnitude of the field strength is often more relevant than small differences in its precise value. For example, measurements of 20~MG and 25~MG differ by 25\%, yet both clearly correspond to the same magnetic field regime and would generally lead to the same astrophysical interpretation. This is well represented by a logarithmic-scale plot, where the order of magnitude prevails. You can see Fig.~\ref{fig:BxB}.

We note that the number of downloaded spectra exceeds that of our initial sample by 25. This is because, in addition to the objects selected from \cite{GF19}, we also retrieved DESI spectra for all magnetic white dwarfs previously identified in the literature. Not all of these stars are included in the \cite{GF19} catalog, nor do they necessarily satisfy our signal-to-noise ratio threshold. 

\begin{figure}
    \centering
    \includegraphics[width=0.45\linewidth]{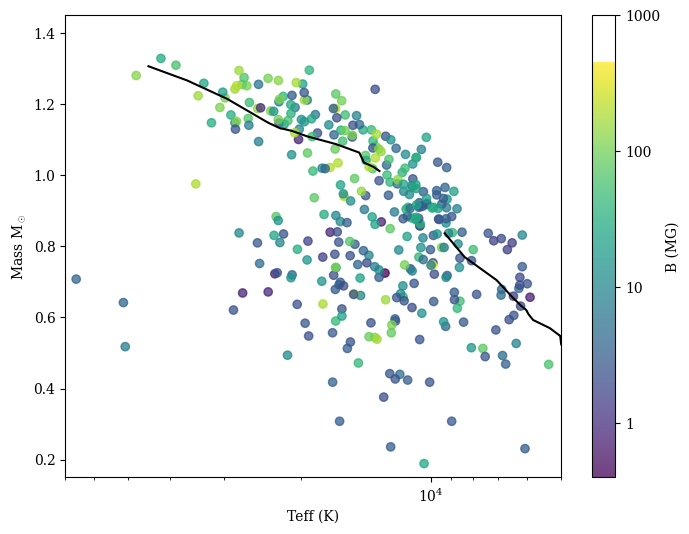}
    \includegraphics[width=0.45\linewidth]{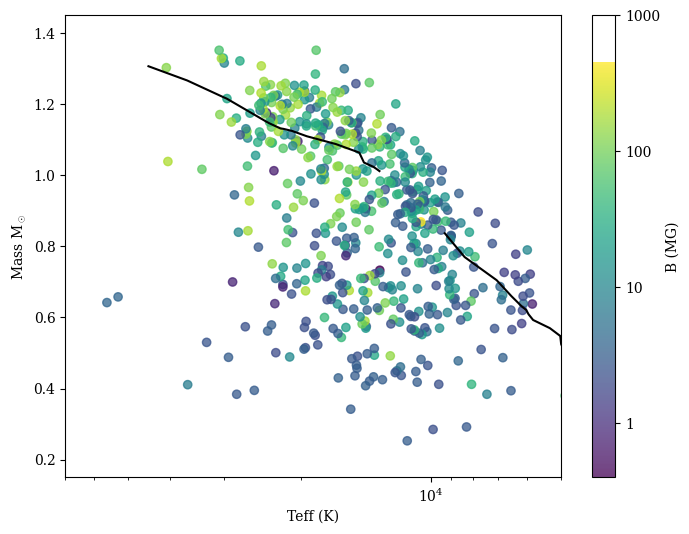}
    \caption{Mass as a function of effective temperature, highlighting the onset of the crystallization sequence. Atmospheric parameters derived from Gaia low-resolution $BP-RP$ spectroscopy and from Gaia photometry are shown in the left and right panels, respectively. 
    }
    \label{fig:cryst}
\end{figure}

To investigate how stellar properties relate to magnetic field strength, we adopt the atmospheric parameters derived by \cite{2024Vincent}. These parameters are based on Gaia $BP-RP$ spectroscopic information and are therefore expected to be more reliable than estimates derived from photometry alone. However, fewer objects have both data available. We compare this in Fig.~\ref{fig:cryst} and see that changes in the numbers or precision do not affect the physical interpretation in this specific case.

One of the main topics in the current discussion of magnetic white dwarfs concerns the possible connection between magnetism and crystallization. As shown in Fig.~\ref{fig:cryst}, the majority of the most magnetic white dwarfs in our sample are concentrated within the crystallization region. However, a significant fraction of the stars with the weakest magnetic fields are more broadly distributed across the diagram and do not appear to be preferentially associated with the crystallization sequence.

\begin{figure}[h]
    \centering
    \includegraphics[width=0.5\linewidth]{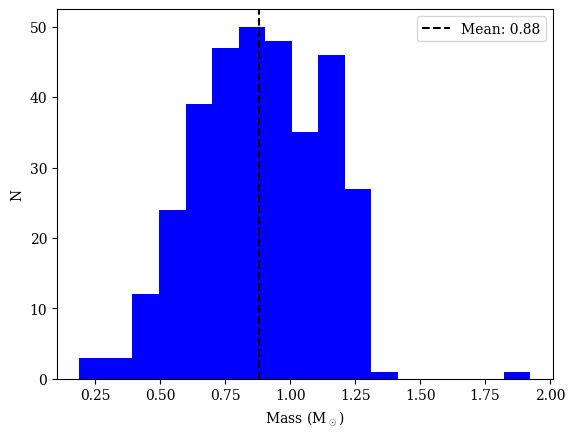}
    \caption{Mass distribution of the magnetic white dwarfs with fields measured from DESI spectra.}
    \label{fig:histmassDAH}
\end{figure}
Numerous studies have shown that magnetic white dwarfs tend to be more massive than their non-magnetic counterparts (see, for example, \cite{2015Ferrario,2013Kepler}). Our sample follows the same trend. Fig.~\ref{fig:histmassDAH} presents the mass distribution of the magnetic white dwarfs in our survey, yielding a mean mass of $M=0.88\,M_\odot$. This value is significantly higher than the mean mass of the overall white dwarf population, further supporting the well-established correlation between magnetism and higher stellar masses. 
We point out that detecting the Zeeman Effect compromises the usual mass determination, as it broadens spectral features and alters the measurement of surface gravity.

\section{Conclusions}

In this work, we present the spectral classification of 44\,417 DESI DR1 white dwarfs, including 11\,685 objects not previously classified in the literature. In this paper, there are 35512 DAs, 2598 DBs, 168 DOs, 3932 DCs, 1021 DZs, and 219 CVs; i.e., DAs correspond to 79,95\% of the white dwarfs with DESI spectra, excluding subdwarfs and CVs. We also derived atmospheric parameters for 29\,072 DA white dwarfs by fitting spectroscopic models to 33\,418 DA spectra, excluding objects outside the applicability of our models or with insufficient signal-to-noise ratios to reliably determine parameters. Using evolutionary models, we derived stellar masses and found a non-Gaussian mass distribution with a mean mass of $0.677\,M_\odot$ and median $0.647\,M_\odot$, consistent with previous large surveys such as SDSS. The distribution exhibits the well-known high-mass component reported in earlier studies.

Moreover, we compiled the largest currently available sample of DESI magnetic white dwarfs with magnetic field strength determinations, identifying 547 magnetic white dwarfs, including 84 new discoveries. Adopting atmospheric parameters from the literature, we find that magnetic white dwarfs have a mean mass of $0.88\,M_\odot$, reinforcing the established result that they are systematically more massive than the general white dwarf population.

Finally, we investigated the distribution of magnetic white dwarfs in the mass--effective temperature plane. While many magnetic white dwarfs are found within the crystallization region, a significant number of stars with intermediate magnetic field strengths lie well before the onset of core crystallization. This result indicates that crystallization-driven dynamos cannot be the sole mechanism responsible for generating magnetic fields in white dwarfs and supports the existence of additional formation channels.

\section{Acknowledgments}
This study was financed in part by the Coordenação de Aperfeiçoamento de Pessoal de Nível Superior – Brasil (CAPES) – Finance Code 001. K.S.O. acknowledges support from the Conselho Nacional de Desenvolvimento Científico e Tecnológico (CNPq).

\bibliography{Reference}{}
\bibliographystyle{aasjournalv7}




\end{document}